\documentclass[fleqn,usenatbib]{mnras}

\usepackage{newtxtext,newtxmath}

\usepackage[T1]{fontenc}

\DeclareRobustCommand{\VAN}[3]{#2}
\let\VANthebibliography\thebibliography
\def\thebibliography{\DeclareRobustCommand{\VAN}[3]{##3}\VANthebibliography}

\usepackage{xr}
\usepackage{xcolor}
\usepackage{graphicx}	
\usepackage{amsmath}	

\usepackage{verbatim}
\usepackage{breqn}
\usepackage[export]{adjustbox}
\usepackage{booktabs}
\usepackage{pgfplotstable}
\usepackage{longtable}
 \usepackage{threeparttable}
\usepackage{pgfplots} 
\pgfplotsset{compat=1.18}

\newcommand{\lya}{Ly$\alpha$}

\newcommand{\kms}{$km s^{-1}$}

\newcommand{\HI}{\mbox{H\,{\sc i}}}

\newcommand{\OVI}{\mbox{O\,{\sc vi}}}

\newcommand{\CIV}{\mbox{C\,{\sc iv}}}

\newcommand{\SiIV}{\mbox{Si\,{\sc iv}}}
   
\title[Low-z metal absorbers]{The role of the ionizing background on the thermal and non-thermal broadening inferred for the low-z intergalactic \OVI\ absorbers}
\author[Mallik et al.]{Sukanya Mallik$^{1}$\thanks{E-mail: sukanyam@iucaa.in}\&
	Raghunathan Srianand$^{1}$
	\\
	\\
	$^{1}$ IUCAA, Postbag 4, Ganeshkhind, Pune - 411007, India}

\date{Accepted XXX. Received YYY; in original form ZZZ}

\pubyear{2022}

\begin{document}

\label{firstpage}
\pagerange{\pageref{firstpage}--\pageref{lastpage}}
\maketitle

\begin{abstract}
Using cosmological hydrodynamical simulations at $z\sim0.5$, we measure the thermal ($b_{t}$) and non-thermal ($b_{nt}$) contribution to the line broadening for the intergalactic absorbers having  \OVI\ and \HI\ absorption well aligned in the velocity space. We find that the inferred temperature based on $b_{t}$ correlates strongly with the optical depth-weighted kinetic temperature of the absorbing gas, albeit with a large scatter. We show this scatter comes from the spread in the kinetic temperature of the gas contributing to the absorption and hence depends on the feedback processes and the ionizing UV background (UVB) used in the simulations.  We show the distribution of $b_{nt}$ is also affected by both feedback processes and the ionizing UVB. 
Therefore, $b_{nt}$ derived using aligned absorbers may not be a good discriminator between the effect of microscopic turbulence and UVB. Instead, the distribution of $b_{t}$ and $b_{nt}$ together with the frequency of occurrence of the aligned absorbers can be used to place additional constraints on the parameters of the simulation for a given assumed UVB.
%
%
\end{abstract}




\section{Introduction}
\label{Sec:introduction}
The \lya\ and metal absorption lines detected in the spectra of distant quasars are frequently used to probe the prevailing physical conditions in the intervening medium. Heavier elements (metals) are believed to originate in stars and are transported into the low-density inter-galactic and circum-galactic medium (IGM and CGM, respectively) via different feedback processes. The detectability of the metals depends on their covering fraction and the thermal and ionization state of the absorbing medium, which are also governed by the spectra of the ionizing Ultraviolet background (UVB). Different feedback mechanisms, such as kinetic and thermal contribution due to stellar and AGN feedback, affect the metal distribution and the physical conditions in the intervening medium. Many cosmological hydrodynamical simulations incorporating a wide range of feedback prescriptions \citep[see for example,][]{dave1999, oppenheimer2009, tepper2011,oppenheimer2012,  tepper2013, rahmani2016, Nelson2018, bradley2022, Mallik2023, khaire2023} are used to understand the influence of feedback processes on statistics of the \lya\ and metal absorbers like column density distribution function (hereafter CDDF), Doppler parameter distribution, line-width distribution and association of metal absorbers with \lya\ absorbers.

The spectra of the  UVB at high redshifts cannot be directly observed and are computed using the spectral energy distribution (SED) and volume densities of quasars and galaxies after taking into account the cosmological radiative transport \citep[refer to][]{haardt1996, faucher2009, haardt2012, khaire2019, faucher2020}. The UVB spectra computed in the literature are uncertain in the extreme UV to soft X-ray energies because of the absence of direct measurements of quasar SEDs due to attenuation by the Milky Way's interstellar medium. The effect of uncertainties in the UVB on the estimated physical parameters like density, temperature, metallicity, elemental abundance, and ionization state of the absorbers has been demonstrated in many previous works, \citep[see for example,][]{schaye2003,simcoe2004,aguirre2004,aguirre2008,howk2009,simcoe2011, fechner2011, hussain2017, Haislmaier2021, Acharya2022}. Very few works like \citet{oppenheimer2009} and recently \citet{appleby2021} and \citet{Mallik2023} discuss the effect of uncertainties in the UVB on the statistics of observed and simulated metal absorbers. \citet{Mallik2023} have shown that the CDDF, cumulative distribution function (CDF) of Doppler parameter (b) and system width ($\Delta V_{90}$), and metal association statistics of \lya\ are affected by the feedback models as well as the UVB. More importantly, it is found that the effect of uncertainty in the UVB on different absorber statistics depends on the feedback used in the simulation, indicating a possible degeneracy in constraining the feedback mechanisms using these statistics. 

The kinetic feedback due to SNe explosions not only influences the metal enrichment in lower-density regions but is also a major source of turbulence in the interstellar and intergalactic medium \citep[]{maclow2004, evoli2011}. While stellar feedback (star formation, stellar wind, and SNe explosion) dominates the turbulent energy input at the scale of the supernova-remnant (10-100~pc) \citep[see for example][]{mckee1977, norman1996, dib2006, avillez2007, joung2009, lu2020}, intermediate and large-scale turbulence also originates due to gravitational \citep{schaye2004, fensh2023, forbes2023} and magneto-rotational instabilities \citep{fleck1981, shellwood1999, beck2015}. At small scales ($<$1~pc), plasma instabilities driven by cosmic rays lead to cosmic ray acceleration and magnetic field amplification. One of the manifestations of the resulting turbulent motion is an additional source of line broadening apart from the thermal contribution \citep{amstrong1995, chepurnov2010, lazarian2023}. Such small-scale turbulence is usually not captured well in cosmological simulations due to insufficient spatial resolution. 

It is a common practice to measure the non-thermal contribution to the line broadening by separating the thermal and non-thermal parts in the Doppler parameter of the aligned absorption of two different species having widely different masses. For high redshift absorbers, \citet{Rauch1996} (using aligned \CIV\ - \SiIV\ lines), \citet{muzahid2012} (using aligned \OVI\ - \lya\ lines) have reported low values of the non-thermal Doppler parameters (with median value $<$ 10 km s$^{-1}$). \citet{Churchill2003} reported turbulent line broadening of $\sim $1 km s$^{-1}$ using aligned Mg~{\sc ii} and Fe~{\sc ii} absorption components for $0.4 \le z \le 1.2$. In contrast, \citet{tripp2008}, \citet{savage2014} have reported (using aligned \OVI\ - \lya\ lines) a much higher value of non-thermal broadening with a median value of 20-26 km s$^{-1}$ for $z<0.5$ absorbers. 
Some previous works \citep[see for example,][]{cen2011, tepper2012, pessa2018} find a trend of higher non-thermal contribution in the \HI\ high column density absorbers.


It has been found that the width of \lya\ absorption in simulations is, on average, narrower than what has been found in observations
\citep[]{meiksin2001, viel2017, Nasir2017}. This is sometimes compensated by introducing turbulent broadening as a subgrid physics in simulations \citep[]{oppenheimer2009, turner2016, gaikwad2017b, maitra2022lowz, bolton2022}. The non-thermal line broadening is also introduced by Hubble flow, local kinematics, and the approximation of blended lines with single/multiple Voigt profiles, etc. The effect of Hubble flow may not be a dominant source as the length of the absorbing medium is, at most, a few hundred kpcs \citep[as discussed for low redshift absorbers in][]{Richter2006, prause2007, hussain2017, mohapatra2021}. The higher non-thermal broadening in the multiphase system reported in \citet{tripp2008} indicates that fluctuations in density-temperature-peculiar velocity fields of the absorbing media can create such effects.


In the simulations, absorption at a given redshift can originate from spatially well-separated regions \citep{peeples2019, Marra2021, Mallik2023} with consistent peculiar velocities. The gradients in the density-temperature field and local kinematics and the decomposition of the absorption profile into single/multiple Voigt profiles can introduce non-thermal line broadening in the simulated absorbers. \citet{Mallik2023} found variations in the UVB alter the regions contributing to the absorption, and the effect depends on the feedback models incorporated in the simulation. In this work, we ask, if we analyse the simulated spectra using Voigt profile analysis (as in observations), how reliably we can derive the thermal and non-thermal contributions to the line broadening.
In addition, we explore the effect of uncertainty in the UVB on the thermal and non-thermal broadening of the aligned \OVI\ and \lya\ absorbers in two sets of cosmological hydrodynamical simulations with different implementations of feedback mechanisms.

While any set of absorption lines originating from atoms with different masses can be used for this exercise, most of the measurements on non-thermal widths at low-z are based on \OVI\ absorbers \citep[see, for example,][]{tripp2008, thom2008, savage2014}. 
Due to the higher relative abundance of oxygen and ionized state, \OVI\ is an important tracer of the low-density, relatively cool photoionized gas and the warm-hot collisionally ionized gas. 
In addition, kinetic temperature measurements in \OVI\ bearing gas are important in estimating their contribution to the cosmic baryon budget.
Therefore,  we focus on low-z aligned absorbers based on \OVI\ and \lya\ absorption. 

This paper is arranged as follows. In section~\ref{Sec:simlations}, we present a summary of the simulations and different ionizing backgrounds used in this work, along with the overview of simulated spectra generation and Voigt profile fitting of the simulated spectra. We also provide details about the method followed to identify the aligned \lya\ and \OVI\ absorbers. We study various statistical distributions of thermal and non-thermal line broadening and their dependence on other physical parameters in section~\ref{Sec:results}. We discuss our results in section~\ref{Sec:discussions}.
\begin{figure*}
	\begin{minipage}{\textwidth}
		\includegraphics[width=\textwidth]{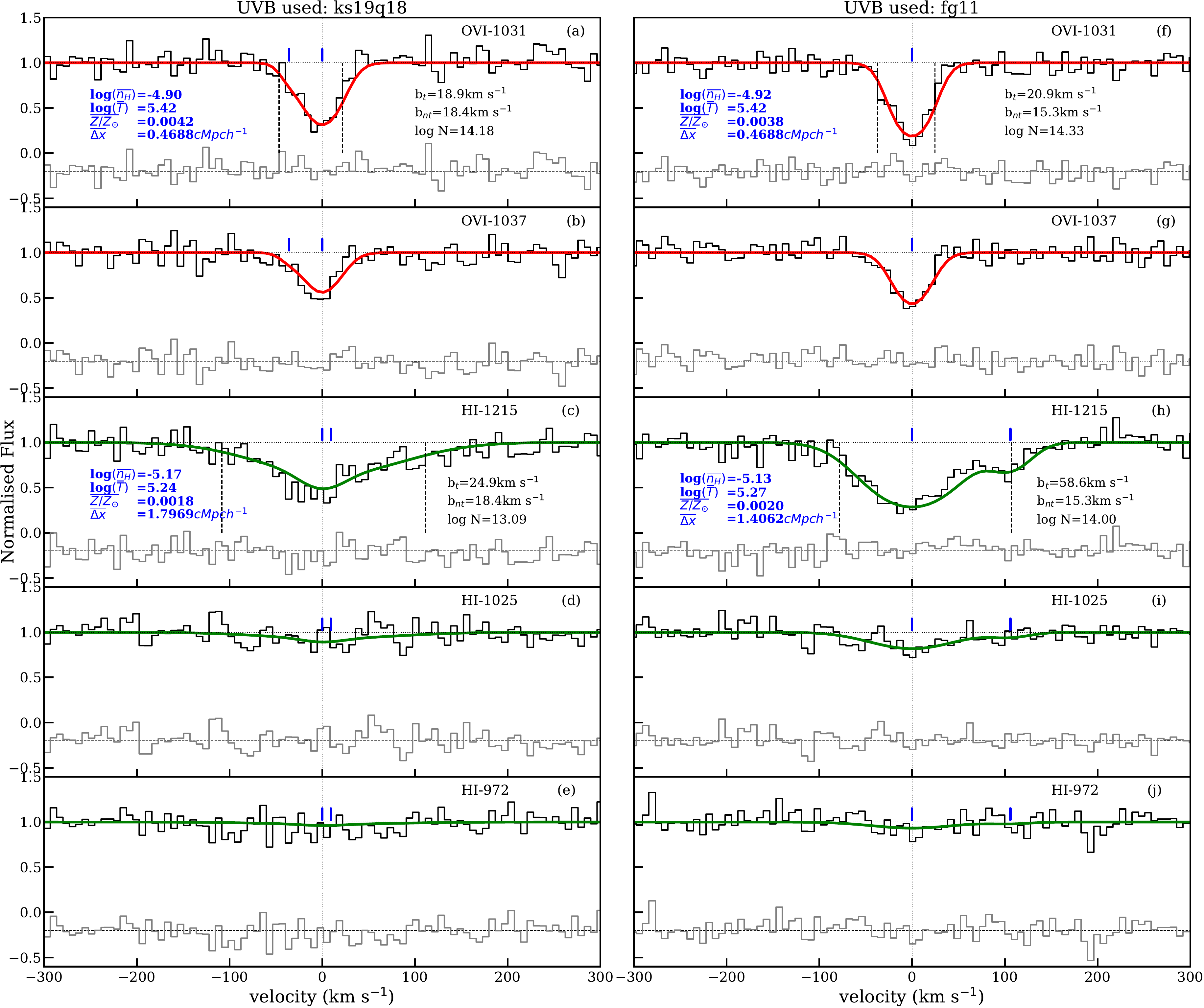}
		\caption{  An example of the simulated spectra (black) of \OVI\ and \lya\ absorber aligned in redshift space from Sherwood "WIND+AGN" simulations.
  The left and right columns show the velocity plots for spectra obtained using "ks19q18" and "fg11" UVBs, respectively. The Voigt profile fits of the aligned \OVI\ - \lya\ absorbers are performed simultaneously using the {\sc vpfit} (green).
The thermal and non-thermal b-parameters and the column density (in log unit) obtained from the simultaneous fit to \OVI\ and \HI\ are indicated in panels (a) and (f) for \OVI\ and in panels (c) and (d) for \lya. The vertical dashed lines in these panels indicate the region used to calculate the system width ($\Delta V_{90}$). We mention different properties of the absorbing gas, e.g., average density ($\overline{n_{H}}$), temperature ($\overline{T}$), metallicity ($\overline{Z/Z_{\odot}}$), and the length of the contributing region $\overline{\Delta x}$ in these panels. Differences seen between the left and right panels illustrate the influence of the UVB on the component structure and the Voigt parameters of \OVI\ and \lya\ absorbers.}
		\label{fig:demo_los}
	\end{minipage}%
\end{figure*}  
\begin{figure*}
	\begin{minipage}{\textwidth}
		\includegraphics[width=\textwidth]{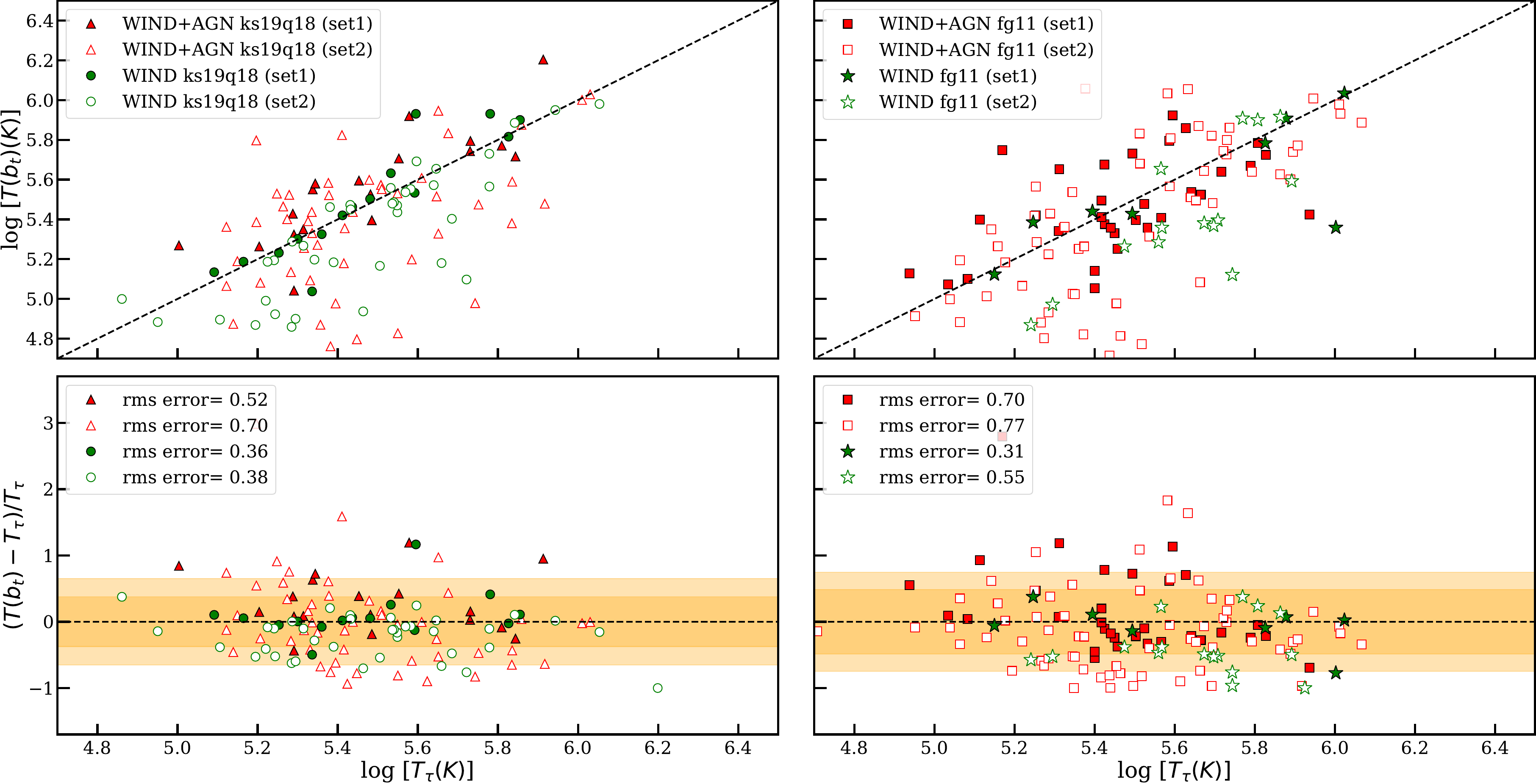}
		\caption{\textit{Upper panels:} Comparison of the temperature estimated from the thermal part of Doppler parameter (b$_{t}$) and optical depth weighted temperature for models using "ks19q18" (left panel) and "fg11" (right panel) UVBs. The dashed line is the equality line.
        The absorbers in "set1" and "set2" (as defined in section~\ref{sec:T_est}) are shown in filled and open symbols, respectively. \textit{Lower panels:} The fractional deviation between the temperatures of the absorbers estimated from two different methods are shown for "ks19q18" (left panel) and "fg11" (right panel) UVBs. The RMS in the fractional deviation of measured temperature with respect to $T_
        \tau$ for the "WIND" and the "WIND+AGN" models are indicated in light and dark-yellow shaded regions, respectively. 
		}
		\label{fig:tau_wt_T_vs_b_therm}
	\end{minipage}%
\end{figure*}
\section{Details of simulations used and different ionizing backgrounds}
\label{Sec:simlations}
In this work, we use the Sherwood simulations \citep[]{Bolton2017} based on parallel Tree-PM SPH code P-Gadget-3, incorporating two different feedback models. The comparison between these two simulations is discussed in detail in \citet[]{Mallik2023} (see section 2 in \citet[]{Mallik2023}). Briefly, both the simulations have boxsize of 80~h$^{-1}$cMpc containing 2$\times$512$^3$ dark matter and baryonic particles and use cosmological parameters \{$\Omega_m$, $\Omega_b$, $\Omega_\Lambda$, $\sigma_8$, $n_s$, $h$\} = \{0.308, 0.0482, 0.692, 0.829, 0.961, 0.678\} from \citet[]{planck2014}. The initial conditions were generated on a  regular grid using the N-\textsc{GEN}IC code \citep[]{springel2005} and transfer functions generated by CAMB \citep[]{lewis2020} at $z=99$. Initial particle mass in Sherwood simulation are $2.75 \times 10^8 h^{-1} M_{\odot}$(DM), $5.1 \times 10^7 h^{-1} M_{\odot}$(baryon) with gravitational softening length set to 1/25$^{th}$ of the mean inter-particle spacing. 
We use two models of the Sherwood simulation suit: the Sherwood "WIND" model, which has only stellar wind feedback, and the Sherwood "WIND+AGN" model, which includes both the wind feedback and AGN feedback. These two models were initiated with the same seed, and all other parameters were the same except for the feedback prescriptions. {The feedback parameters implemented in the Sherwood simulations were chosen such that the galaxy properties like galaxy stellar mass function and redshift evolution of star formation rate density are consistent with the observations \citep[see section 3.1-3.3 in][]{Puchwein2013}. The wind and AGN feedback models used here broadly agree with the other recent simulations like EAGLE \citep[]{schaye2015, crain2015} and Illustris TNG \citep[]{pillepich2018, weinberger2017}, but are less severe than the AGN feedback used in SIMBA simulation \citep[]{dave2019}. 
\citet{Puchwein2013} found the convergence for the wind feedback-only models is well established, but the models, including both wind and AGN feedback, may have some convergence issues in their lowest resolution model (with mass resolution 2.75 times higher than the model used in this work). This may not be a major issue for our study, as we aim to probe how well the temperatures derived from the aligned absorbers trace the underlying gas temperature.
 The global metallicity of the gas particles is retained, and we assume solar relative abundance of different elements.} 
 The volume, as well as the resolution of the simulation box and the implemented feedback prescriptions, are comparable to the state-of-the-art simulations used to study metal absorbers in IGM \citep[for example,][]{oppenheimer2009, oppenheimer2012, tepper2011, tepper2013, Nelson2018}.


Both simulations used here were run considering the UVB computed by 
\citet{haardt2012}. In order to explore the effect of changing the UVB
we consider the UVBs computed by \citet{khaire2019} (ks19q18; corresponding to a far UV spectral index $\alpha=-1.8$) and by \citet{faucher2009} (the 2011 updated version of it, hereafter referred to as "fg11") while generating the spectra. The detailed comparison of different available UVBs can be found in section 2.1 of \citet{Mallik2023}.  While using these UVBs we make an explicit assumption that the effect of variation in the UVB on the gas temperature is negligible and mainly consider its effect on the ionization state of the absorbing gas. This assumption is reasonable as the gas temperature at low redshift ($z \leqslant 0.5$) IGM is dominated by the adiabatic expansion cooling. The \OVI\ photoionization rate in the case of ks19q18 UVB is 2.83 times higher than that of the fg11 UVB, and these two UVBs span the range of \OVI\ photoionization rate reported in the literature. 

\subsection{Generating simulated metal spectra}
We shoot lines of sight through the simulation box and generate transmitted flux spectra. We calculate the number-density of a given ion species from the density and metallicity of SPH particles considering the ionization correction obtained from photoionization code~{\sc cloudy} for the assumed UVB. For simplicity, we consider the optically thin conditions while computing the ion fractions. This is a valid assumption as we found later that most of our \lya\ absorbers have \HI\ column density lower than the threshold \HI\ column density \citep[given by][]{rahmati2013} where self-shielding becomes important. We divide the sightline (80~h$^{-1}$cMpc) into 1024 equispaced grids ($\sim$7 km s$^{-1}$), and at each grid point, we assign ion number density (n$_{XI}$), temperature (T), and peculiar velocity (v) field of 
following standard SPH formalism described in \citet{monaghan1992} and \citet{springel2005a}.
 We use these quantities in the grid points along the line of sight to calculate the optical depth of a given ion species ($\tau_{XI}$), following the sightline generation procedure described in section 2.3 in \citet{Mallik2023}. We convolve the transmitted flux (F$_{XI}$= exp(${-\tau_{XI}}$)) with a Gaussian profile with a FWHM of  17 kms$^{-1}$, mimicking the instrumental resolution of HST/COS spectrograph and added Gaussian noise corresponding to SNR=10 per pixel. 

\subsection{Selection criteria and Voigt profile fitting for the aligned absorbers}
We have generated 10000 sightlines following the procedure described above. The initial identification of absorption lines and Voigt profile fitting were performed using the automated Voigt profile fitting code, {\sc viper} \citep[see][for details]{gaikwad2017b}. Currently, {\sc viper} can automatically identify absorption systems of individual ions above a chosen significant level (i.e., Rigorous Significant Level as defined by \citet{keeney2012}, RSL $\ge$ 4, in our case) and fit multiple Voigt profile components minimizing $\chi^2$ and Akaike information criterion(AIC).
However, simultaneous fitting of multiple transitions from different ions from a single system (that is required to decompose the thermal and non-thermal contribution to the line broadening) is at present not possible with {\sc viper}. For this purpose, we use the standard Voigt profile fitting code {\sc vpfit} \citep[version 10.4,][]{Carswell2014} frequently used by the observers. 
%

We initially use {\sc viper} to independently identify and fit all the \OVI$\lambda 1032$ and \lya\ absorption lines (having RSL$\ge$ 4) along the 10000 spectra generated from our simulations. 
We then make a subset absorbers producing both \OVI\ and \lya\ absorption with their
line centers separated by $\leqslant$ 10 kms$^{-1}$. We refer to them as aligned absorber candidates. Previous works both in observation \citep[see for example][]{Rauch1996, tripp2008, muzahid2012, savage2014, sankar2020} and simulations \citep[such as][]{oppenheimer2009, tepper2012, pessa2018} consider \OVI\ and \lya\ components either having same redshift or a velocity offset within the error of the velocity-centroid measurement as the aligned absorbers. Thus, our identification of an aligned absorber is conservative compared to those definitions.
%

We then performed simultaneous multi-component Voigt profile fitting of all the detectable \HI\ Lyman-series absorption lines and the \OVI\ doublet lines of the candidate aligned absorbers using the {\sc vpfit}. During the fit, we constrained the redshift to be identical (i.e., complete alignment for \HI\ and \OVI\ components). For this, we used the `tied parameters' option in {\sc vpfit} for redshifts and b-parameters of \OVI\ and \lya\ components. The column densities of the ions are freely varied during the Voigt profile fitting. The fitting procedure also returns the best-fitted temperature and the non-thermal contribution to the Doppler parameter using the tied ions.  
We obtained the best-fitted values of column densities ($N_{\rm OVI}$ and $N_{\rm HI}$) and the Doppler parameter due to thermal (b$_{t}$) and non-thermal (b$_{nt}$) broadening of the line for individual components.


%

As an example,  in Figure~\ref{fig:demo_los}, we show the simultaneous Voigt profile fitting results to \HI\ and \OVI\  absorption in one of the aligned absorbers from "WIND+AGN" simulations when we consider the "ks19q18" (left panels) and "fg11" (right panels) UVBs.
%
The mean density, temperature, metallicity, and the length of the gas-region contributing to the absorption are influenced by the UVB, as indicated by the legends provided in Figure~\ref{fig:demo_los}. The differences in the physical properties of the contributing region due to variation in the UVB results in significant differences in the component-structure of the absorption lines, as illustrated for both \OVI\ and \lya\ absorption profiles in the Figure~\ref{fig:demo_los}. This leads to slightly
different best-fitted Voigt profile parameters, as can be seen from the legends in Figure~\ref{fig:demo_los}. 

\begin{figure*}
	\begin{minipage}{\textwidth}
		\includegraphics[width=\textwidth]{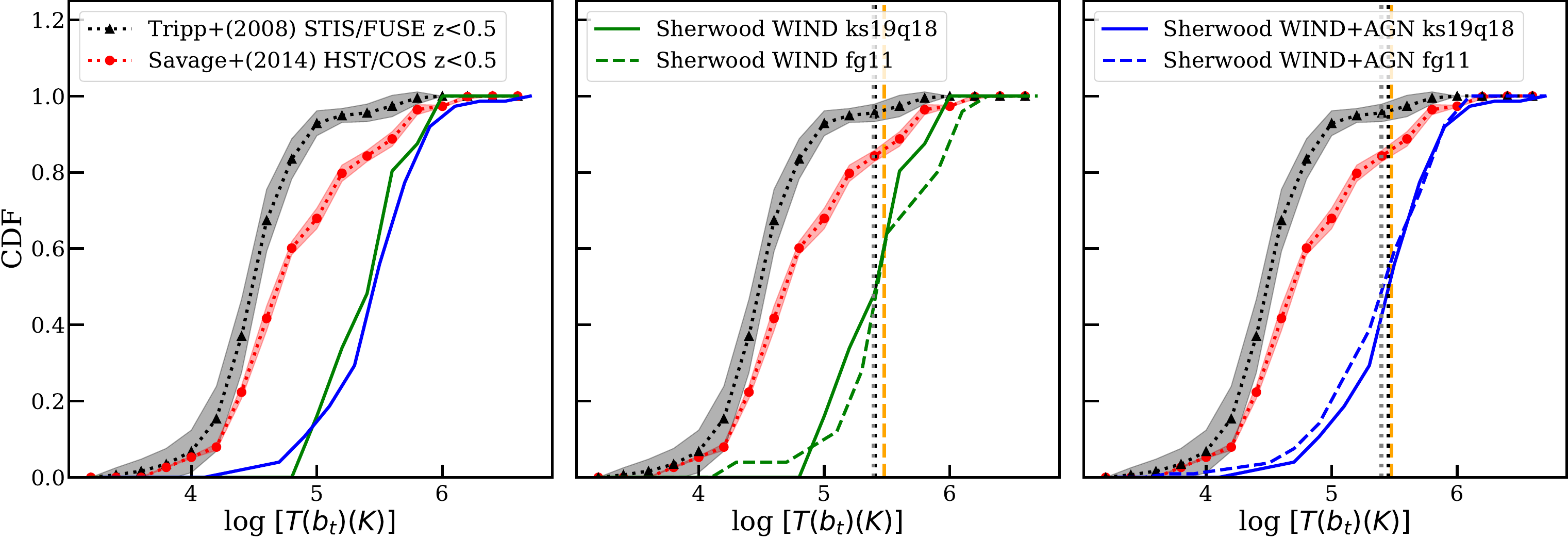}
		\caption{The cumulative distribution function for temperature estimated from b$_{t}$ with simulated spectra obtained from Sherwood "WIND" and "WIND+AGN" using UVB "ks19q18" are shown in the left panel. Following the procedure described in section~\ref{subsec:Tbt_dist}, the observed $T(b_{t})$-CDF and corresponding errors from \citet{tripp2008} (black) and \citet{savage2014} (red), shown in dotted line and shaded region respectively. The comparison between $T(b_{t})$ between spectra obtained using UVB "ks19q18" and "fg11" is shown for the "WIND" model in the middle panel and for the "WIND+AGN" model in the right panel. The vertical dotted lines indicating the median $T(b_{t})$ for the "WIND" (grey) and the "WIND+AGN" (black) model are close to the temperature of peak fraction in collisionally ionized \OVI\ (T$\sim 3 \times 10^{5}$K), shown in orange dashed line. }
		\label{fig:T_cdf}
	\end{minipage}%
\end{figure*}

\begin{figure*}
	\begin{minipage}{\textwidth}
		\includegraphics[width=\textwidth]{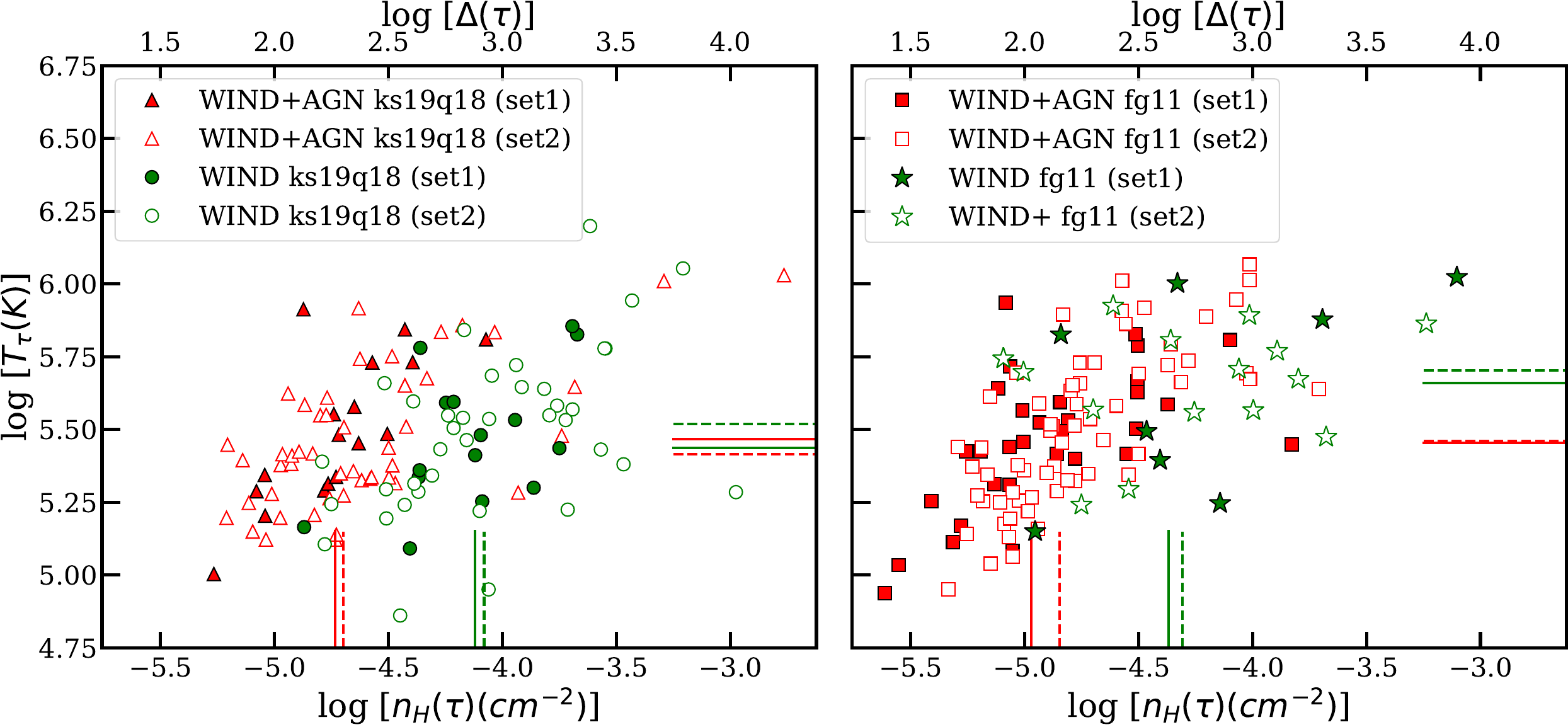}
		\caption{The optical depth weighted temperature ($T_{\tau}$) and density ($n_H(\tau)$) of the aligned absorbers in the "WIND" and "WIND+AGN" simulations for "ks19q18" UVB is shown in the left panel. The absorbers in "set1" and "set2" (as defined in section~\ref{sec:T_est}) are shown in filled and open symbols. The vertical and horizontal lines (solid line for "set1" and dashed line for "set2") indicate the median density and temperature, respectively. A similar plot for "fg11" UVB is shown in the right panel.}
		\label{fig:tauT_tauNH}
	\end{minipage}%
\end{figure*}
\section{Results}
\label{Sec:results}
%
It has been shown in \citet{Mallik2023} that the b-distribution is significantly affected by the feedback processes (see their Figure 8). Here, by decomposing the b into thermal (b$_t$) and non-thermal (b$_{nt}$) parts, we study the impact of feedback effects on the inferred thermal and the non-thermal line broadening. We also study how these results are affected by the assumed UVB.

\subsection{Temperature estimation of the absorbers}
\label{sec:T_est}

 For each of the Voigt profile components obtained using the VPFIT, we calculate the optical-depth-weighted temperature (T$_\tau$) from the simulations using the following procedure. Let $\tau_{ij}$ is the optical depth contribution at the $i^{th}$ pixel from the gas at $j^{th}$ pixel having a temperature $T_j$.
Then the optical depth weighted temperature at the $i^{th}$ pixel $T_\tau(i)$ is given by,
$T_{\tau}(i)=\sum_{j=1}^{N} T_j \tau_{ij}/\sum_{j=1}^{N} \tau_{ij}$.
%
As \lya\ profile often shows multiple components and not all of them produce detectable \OVI\ absorption, we used the \OVI\ optical depth for calculating $T_{\tau}$. 
We then assign the median value of $T_{\tau}(i)$ for pixels within the system-width of the absorption to be the optical-depth-weighted temperature ($T_{\tau}$) of the absorber. Next, we explore the possible connection between the kinetic temperature derived from $b_t$ and $T_\tau$ for individual voigt profile components.

In the left panel of Figure~\ref{fig:tau_wt_T_vs_b_therm}, we compare the kinetic temperature [$T(b_t)$] derived from $b_t$ for the aligned absorption components with $T_\tau$ for Sherwood "WIND" and "WIND+AGN" simulations when we use the "ks19q18" UVB. 
Firstly, we notice a strong correlation between $T_\tau$ and $T(b_t)$ for all the absorbers. This suggests that T($b_t$) is a good representation of the optical depth-weighted temperature of the regions contributing to the absorption. However, there is a large scatter around the equality line. The scatter seems to be higher for the WING+AGN simulations compared to the WIND-only simulations. To quantify this, in the lower panel, we show the fractional deviation between these two estimates of temperature (defined as $(T(b_{t})-T_{\tau})/T_{\tau}$)  for both the simulations. We find the RMS of the scatter to be 0.37 and 0.65 for "WIND" and "WIND+AGN" simulations, respectively.


 To understand the origin of the scatter, we have further divided the absorbers into two sets. The first set ("set1") contains absorbers where both \OVI\ and \lya\ absorption profiles are consistent with a single Voigt component. Hence, the effect introduced by the degeneracy associated with the multiple-component Voigt profile fitting is expected to be minimal. 
The "set2" contains absorbers with \lya\ absorption having multiple components, and only one or few of these components are aligned with  \OVI\ absorption component(s).
In this case, it is possible that some of the derived parameters of \lya\ absorption may be influenced by the uncertainty in the Voigt profile decomposition.

The "WIND" and "WIND+AGN" simulations have 26-29$\%$ absorbers in "set1" for "ks19q18" UVB. The RMS of the fractional deviation in the case of "WIND+AGN" simulations are 0.52 and 0.70, respectively, for "set1" and "set2".  This is in line with our expectations. 
%
In the case of WIND-only simulations, the RMS of the fractional deviation is  0.36 and 0.38 for "set1" and "set2" respectively. Even though there is an increase in the RMS value for "set2", the increase is much less pronounced compared to what has been found for the WIND+AGN simulations.
 To explore this scatter further, in the case of "set2", we regenerated a modified profile of \lya\ absorptions, considering only the pixels that contribute to the \OVI\ absorption of the aligned component. This eliminates the possible source of scatter originating from the uncertainty in assigning a particular \lya\ component as an aligned component in a multi-component blended \lya\ profile. We find the scatter in $T(b_t)$ around the optical depth weighted temperature in the "WIND+AGN" model reduces by $20-25\%$ compared to the full sample of absorbers. However, the scatter (RMS $\sim 0.52-0.63$) is comparable to or slightly less than the scatter found for absorbers in "set1". 
 
\citet{Mallik2023} have shown that the length of the region contributing to a given absorption can vary with the assumed UVB. This means the scatter seen in the fractional deviation could also depend on the assumed UVB.
In the right panel of Figure~\ref{fig:tau_wt_T_vs_b_therm}, we show the results considering the "fg11" UVB. As in the case of "ks19q18" UVB, the inferred  T(b$_{t}$) in the case of "fg11" UVB is broadly consistent with $T_{\tau}$. However, we do see the scatter around the unity line is increased in the case of "fg11" UVB, i.e., the RMS increases to 0.49 and 0.75 for the "WIND" and "WIND+AGN" models, respectively (shown in the yellow shaded regions in the right-lower panel) when we considered the full sample. 
Similar to the case of "ks19q18" UVB, around $30\%$ of the absorbers in both simulations are in "set1" for "fg11" UVB. The multi-component absorbers ("set2") in the case of "fg11" also show a larger scatter from the equality line compared to the single-component absorbers ("set1").  {The RMS of fractional deviation for absorbers in "set1" and "set2" in the "WIND" simulation is 0.31 and 0.55, respectively. The scatter is considerably higher in the case of the "WIND+AGN" simulation, with RMS of fractional deviation being 0.70 and 0.77, respectively. }
%
%
This is easy to understand as
more low-density regions contribute to the absorption when softer UVB like "fg11" is used. 
This increases the range of density (temperature) of the gas contributing to a given absorption, and hence, we expect the T(b$_{t}$) to deviate more from the optical depth weighted value, as reported here. 


%
\textit{Overall the temperature estimated from b$_{t}$ roughly agrees with the optical-depth-weighted value for both "WIND" and "WIND+AGN" simulations for spectra obtained using both "ks19q18" and "fg11" UVBs. 
The scatter in this relationship is influenced by the feedback process, UVB, and degeneracies associated with Voigt profile decomposition when multiple components are present.
}

\begin{figure*}
	\begin{minipage}{\textwidth}
		\includegraphics[width=\textwidth]{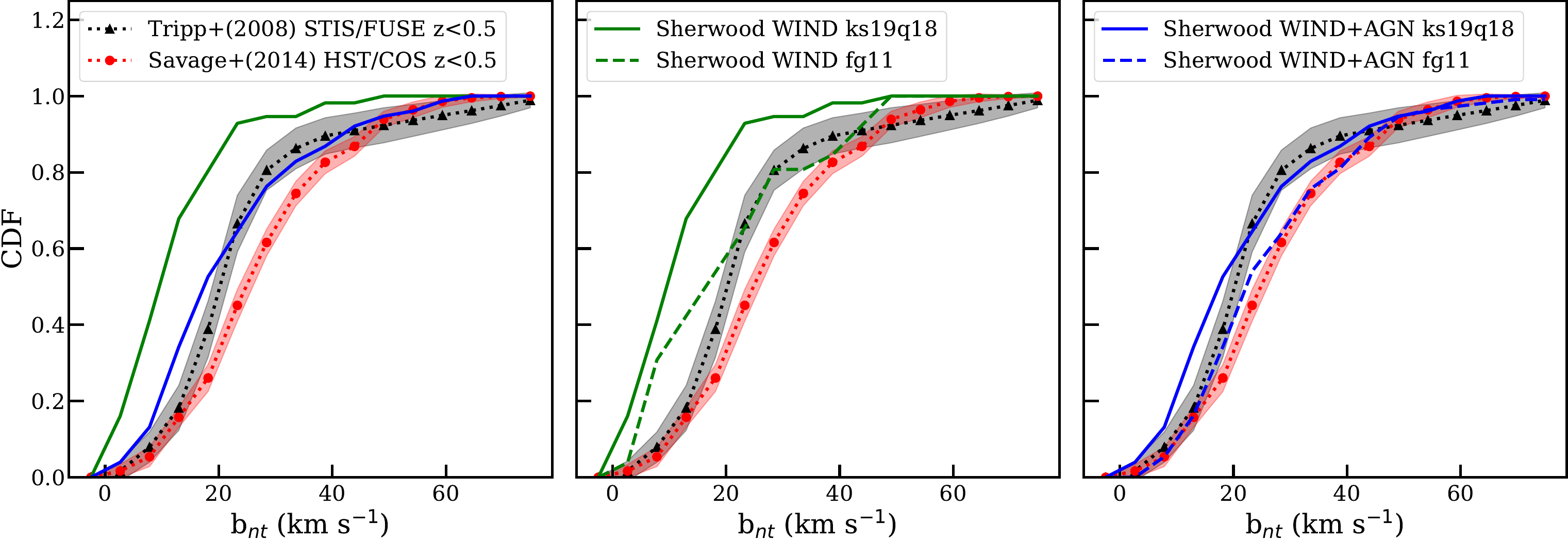}
		\caption{Same as that of Figure~\ref{fig:T_cdf}, but for the non-thermal part of Doppler parameter (b$_{nt}$). The observed data from \citet{tripp2008} and \citet{savage2014} are shown in black and red markers, respectively. 
		}
		\label{fig:bnt_cdf}
	\end{minipage}%
\end{figure*}
\begin{figure*}
	\begin{minipage}{\textwidth}
		\includegraphics[width=\textwidth]{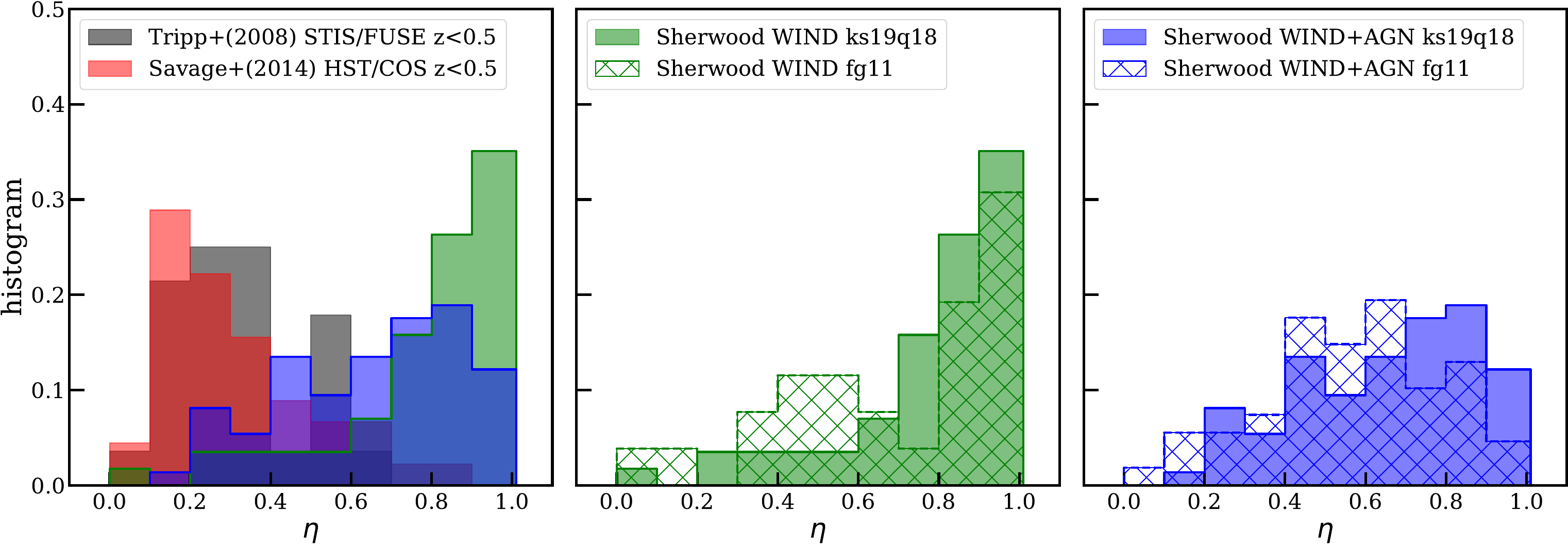}
		\caption{The distribution of parameter $\eta$ in the Sherwood "WIND" and the "WIND+AGN" models for "ks19q18" UVB are shown in the left panel. Both simulations show higher thermal contribution than the observed data from \citet{tripp2008} and \citet{savage2014} indicated in black and red filled histograms, respectively. The effect of using "fg11" UVB instead of "ks19q18" is shown in using hatched histograms in the middle and the right panels for the "WIND" and "WIND+AGN" models, respectively.  
		}
		\label{fig:eta_hist}
	\end{minipage}%
\end{figure*}

\subsection{Temperature distribution of the absorbers}
\label{subsec:Tbt_dist}

From figure 8 of \citet{Mallik2023}, we can see the b-distribution for the  \OVI\ components obtained for the Sherwood simulations for AGN+WIND model roughly follow the observed distribution from \citet{danforth2016} for both the UVBs considered here. In the case of WIND-only models, the b-distribution is closer to the observations when we assume the ``fg11" UVB. Now, we ask whether the temperature derived for the aligned absorbers shows a similar trend.

In Figure~\ref{fig:T_cdf} left panel, we show the distribution of T(b$_{t}$) for simulated spectra using "ks19q18" UVB for "WIND" and "WIND+AGN" models. We generated 500 sets of mock observation values from Gaussian random number selection within the error range of each observed temperature value assigned to each of the aligned \OVI\ - \lya\ pairs and calculated CDF for each set. The mean value of these CDFs is plotted in red/black symbols, and the error in CDF is obtained from the standard deviation of these sets.
We compare the temperature distribution of the \OVI\ absorbers obtained from the simulated spectra with the observed distribution reported in \citet{tripp2008} and \citet{savage2014}. The two observed distributions do not match well. As discussed in \citet{savage2014} the difference between the two sets of data could originate mainly from the systematically higher SNR achieved in the HST/COS observations used in \citet{savage2014}.


 Unlike what has been found for the b-distribution by \citet{Mallik2023},
 the distribution of T($b_t$) from the simulations is systematically higher than what has been found in observations. The difference is much larger than the difference seen between the two measurements from the observed spectra used here.
 The median temperatures in the two simulations ($10^{5.35}$K and $10^{5.43}$K for "WIND" and "WIND+AGN" models, respectively) differ by 20$\%$. However, the temperature distribution of the aligned absorbers in these two simulations is not significantly different, having median T(b$_{t}$) close to the temperature of peak fraction in collisionally ionized \OVI\ (T$\sim 3 \times 10^{5}$K). This is also confirmed by the KS test, resulting in a p-value of 0.20. The temperature distribution of absorbers also does not differ significantly between "set1" and "set2" in both simulations, although the KS-test, in this case, may be affected by the small number of data points. { Just for the completeness sake the observed median value of $T(b_t)$ are  $10^{4.5}$K and $10^{4.6}$K for \citet{tripp2008} and \citet{savage2014} respectively. }

We compare the T(b$_{t}$)-CDF obtained for simulated spectra generated using "ks19q18" uvb with "fg11" in the middle (for the "WIND") model and right (for the "WIND+AGN" model) panel. Both these panels show T(b$_{t}$)-CDF in "fg11" UVB also does not agree with the allowed range from observation. The median value in "fg11" increases by 20$\%$ for the "WIND" model but decreases by 10$\%$ for the "WIND+AGN" models. The p-values in the KS-test for both the feedback models are $>0.05$, which implies T(b$_{t}$) distribution does not change significantly with variation in UVB. This is expected as the temperature of individual SPH particles is not modified when we use different UVBs. 

To further investigate the physical conditions of the aligned absorbers, we compared the absorbers in optical depth weighted temperature-density plane in Figure~\ref{fig:tauT_tauNH}. The left panel shows the absorbers in "ks19q18" UVB. Although the absorbers in both simulations trace temperatures in the collisional ionization range, the majority of absorbers in the WIND-only model trace higher density medium compared to the absorbers in the "WIND+AGN" simulation. This behavior is commensurate with the finding in \citet{Mallik2023} that the absorbers found in the  "WIND" model trace higher-density regions compared to those from the "WIND+AGN" model. The absorbers for "fg11" UVB in two simulations also show a similar trend, as shown in the right panel of Figure~\ref{fig:tauT_tauNH}. These absorbers also trace lower-density regions compared to "ks19q18" UVB, again confirming the conclusion in \citet{Mallik2023} that more absorbers can originate in lower-density regions for softer UVB. 

\subsection{Distribution of non-thermal b-parameters}

We show the Cumulative distribution of b$_{nt}$ for aligned b(\OVI)-b(\HI) pairs in Figure~\ref{fig:bnt_cdf}. In the same figure, we compare our results with the observations reported in \citet{tripp2008} and \citet{savage2014}, plotted in black and red points. The b$_{nt}$-CDF and corresponding errors from the observation data are calculated in a similar procedure as T(b$_{t}$)-CDF of observation, as described in section~\ref{subsec:Tbt_dist}. The error in b$_{nt}$ is found by propagating the errors in $b$ and $T$. Like in the case of $b_t$ the two observed cumulative distributions for $b_{nt}$ are not consistent with one another. This was attributed to the improved SNR in the case of \citet{savage2014}.

%
%

From the left panel in Figure~\ref{fig:bnt_cdf}, we can clearly see that the cumulative distribution of $b_{nt}$ for the two simulations are different.  This is confirmed by the KS-test with a  p-value of $<$ $10^{-4}$, which suggests  $b_{nt}$ from the "WIND" and "WIND+AGN" models are not tracing the same population.
The median b$_{nt}$ of the "WIND+AGN" model (=16.06 km s$^{-1}$) is 1.8 times higher than the "WIND" model (=9.09 km s$^{-1}$) for "ks19q18" UVB.
This is in line with our expectation that the inferred $b_{nt}$ will be influenced by the feedback processes.


The middle and right panels of Figure~\ref{fig:bnt_cdf}, we study the effect of UVB on the derived $b_{nt}$ distribution. First, we notice that for both simulations, the distribution of $b_{nt}$ depends on the UVB used. The $b_{nt}$ values tend to be larger for the "fg11" UVB that has a lesser \OVI\ and \HI\ ionization rate compared to that of "ks19q18" in both simulations. In the case of "WIND+AGN" simulations the median $b_{nt}$ value increases from  16.86 \kms\ for "ks19q18" to  21.83 \kms\ for "fg11" UVB. Similarly, in the case of the "WIND" only model, the median $b_{nt}$ value increases from  9.09 \kms\ for "ks19q18" to  15.84 \kms\ for "fg11" UVB.

 Having established the result that the $b_{nt}$ distribution derived for the aligned \OVI-\HI\ absorbers is sensitive to the spatial distribution of ions that are influenced by the feedback processes and assumed UVB, we compare the distributions from the simulations with the observed one.
The observed data from \citet{tripp2008} and \citet{savage2014} has median b$_{nt}$ values of 20 \kms\ and 26 \kms, respectively. In the case of "WIND" only simulations, both the UVBs produce median $b_{nt}$ lower than what has been observed. From the right panel, it is clear that the distribution of $b_{nt}$ is closer to the simulations.
However, in order to have a realistic comparison, we need to simulate the spectra with similar SNR distribution as in the observations and use a more realistic line spread function that is beyond the scope of this work.

%

\subsection{Thermal vs. non-thermal contribution in line-broadening}
\label{sec:eta}
We characterize the nature of line broadening (i.e., whether thermal or non-thermal broadening dominates) of aligned absorbers using a parameter $\eta \equiv \sqrt{\frac{2kT}{m_{ion}b^2}}$, as previously done in \citet{cen2011}. Note that the value of $\eta$ varies between 0 to 1 depending on the relative importance of thermal and non-thermal broadening in the Doppler parameter; for a purely thermal broadened absorber, $\eta=1$ and a non-thermal broadening-dominated absorber has $\eta=0$.

We show the histogram of $\eta$ for the "WIND" and "WIND+AGN" model for "ks19q18" UVB in the left panel of Figure~\ref{fig:eta_hist} along with the observations from \citet{tripp2008} and \citet{savage2014}.
As expected based on the discussions presented till now, most of the absorbers in the "WIND" model have $\eta$ values near 1, implying mostly thermal broadening dominated Doppler width, whereas the "WIND+AGN" model has relatively lower $\eta$ values, implying higher importance of non-thermal broadening. 
This suggests that the distribution of $\eta$ can provide
interesting constraints on the feedback models used in the simulations.
As expected based on the discussions above, both models have lower non-thermal contributions in line broadening compared to the observation, which is also influenced by the fact that the inferred temperatures simulation temperatures are high for the simulations used here. 

The $\eta$ histogram in the middle and right panels shows non-thermal contribution increases when "fg11" UVB is used in both the "WIND" and the "WIND+AGN" simulations, respectively. 
We believe the effect of UVB on the $\eta$ distribution will be higher if the gas temperature is lower compared to what we find in Sherwood simulations.  This is because the contribution of photoionization is expected to be higher when the gas temperature is lower.

Some previous works \citep[see for example,][]{cen2011, tepper2012, pessa2018} reported $\eta$ could depend on the \lya\ column density with high column density absorbers having more non-thermal contribution. The observed data from \citet{savage2014} shows weak anti-correlation with a Pearson correlation coefficient equal to -0.37 (p-value =0.01).
However, 
%
we do not find any significant correlation in the observed data from \citet{tripp2008} or for the simulations used here for both the UVB as confirmed by a low value of the Pearson correlation coefficient ($<0.30$) and/or high p-values ($>0.05$). 

\section{Summary and discussions}
\label{Sec:discussions}
Absorption lines produced by different ion species of the aligned absorbers are often used to probe the thermal and non-thermal contribution to the absorption line-broadening. Such measurements are used to infer the kinetic temperature of the absorbing gas and the non-thermal contribution to the velocity spread. The temperature estimates using this approach is very important to identify the dominant ionizing mechanism (i.e. photo- or collisional ionization). The non-thermal contribution is broadly interpreted as turbulence (or missing sub-grid physics) in the literature.
It is well established through simulations that absorption at a given redshift originates from spatially well-separated regions \citep{peeples2019, Marra2021, Mallik2023} having consistent peculiar velocities. Therefore, it is important to understand what we measure using Voigt profile fitting techniques as we assume the absorption to originate from a single cloud with constant temperature and density.

%

Recently, we have shown that the distribution of Doppler parameters of the simulated \OVI\ absorbers is influenced by the feedback model and the choice of UVB used in the hydrodynamical simulation \citep[]{Mallik2023}. In particular, the spectra generated from the WIND+AGN Sherwood simulation are found to be closely reproducing the observed b distributions of \OVI\ when we use 'fg11" UVB (see their figure 8). This study also emphasized the importance of the simultaneous usage of several statistical distributions to lift the degeneracy between the effect of feedback and the UVB assumed. 

Here, using cosmological simulations, we probe how well the temperature derived using the aligned absorbers probe the underlying gas temperature.
We also probe how the derived non-thermal width is affected by the feedback process and the assumed UVB.
For this, we consider two different simulations (viz., Sherwood "WIND" model and Sherwood "WIND+AGN" model) having varied feedback. We also explore the effects of the background radiation field, using two different UVB models viz. "ks19q18" and "fg11". Similar to the method followed in observation, we identify absorbers with aligned \OVI\ and \lya\ absorption and simultaneously fit them using {\sc vpfit} to find the thermal and non-thermal contribution to the line-broadening. 

 We show the temperature of the absorbers estimated from the thermal contribution in the line-broadening correlates strongly with the underlying optical depth-weighted temperature of the absorbing gas, although showing RMS scatter of 0.37-0.49 and 0.65-0.75 (in the fractional deviation from zero) for the "WIND" and the "WIND+AGN" models, respectively considering two different UVBs. The single-component isolated \OVI-\lya\ aligned absorbers have much less scatter compared to the aligned absorbers in a multi-component blended system.
 We show the scatter primarily comes from the spread in kinetic temperature of the gas contributing to the absorption and hence depends on the feedback processes and the UVB. The uncertainty in the Voigt-profile decomposition further increases the scatter in the case of blended systems. 

We find the median temperature ($T(b_t)$) in the "WIND" model ($\sim 10^{5.35}$K) is 20$\%$ higher than that of the "WIND+AGN" model ($\sim 10^{5.43}$K) for the "ks19q18" UVB. 
We find the median temperature value found in the simulations used here is higher compared to the observed median value of temperature ($< 10^{4.6}$ K) from \citet{tripp2008} and \citet{savage2014}. This confirms some of the shortcomings of Sherwood simulations found in \citet{Mallik2023}.
The median value of non-thermal line broadening of aligned absorbers in the "WIND+AGN" model (= 16.06 km s$^{-1}$) is 1.8 times higher than that in the "WIND" model (= 9.09 km s$^{-1}$). The non-thermal contribution increases in the case of "fg11" UVB.


The relative contribution of thermal broadening in the Doppler parameter of an absorber is estimated using a parameter $\eta$  (as defined in section \ref{sec:eta}). The absorbers in the "WIND" model have $\eta$ values mostly near 1, whereas the absorbers in the "WIND+AGN" model have relatively lower $\eta$ values. The histogram of $\eta$ for simulated absorbers does not agree with the observed distribution, which is partly because of higher temperature estimation in the simulations used here.

{The main caveats in this work are that we assumed the change in UVB only alters the ionization state and not the temperature of the gas, and we have also assumed the gas to be optically thin. We plan to address this issue by using different UVBs during the simulation run in our future work. It will also be an interesting exercise to explore the effects of the feedback model variations on the non-thermal line broadening using a wide range of feedback models in a higher resolution simulation box like Illustris TNG, as varied feedback results in large dispersion in \OVI\ CDDF \citep[see Figure 6 in ][]{Nelson2018}}. 

In conclusion, we find that the derived temperature using the aligned absorber roughly provides the optical depth-weighted kinetic temperature of the gas. The non-thermal broadening inferred arises due to the inhomogeneities in the temperature-density fields and from the uncertainties associated with the multiple-component Voigt profile decomposition. Therefore, the derived non-thermal line-broadening using aligned absorbers may not be probing the sub-grid turbulence arising due to stellar feedback or gravitational instabilities. Together with flux statistics (flux probability distribution function and two-point correlation function) as well as column density and Doppler parameter distribution, the number of metal absorption aligned with a \lya\ component and distribution of thermal and non-thermal line broadening can be used to place additional constraints on the feedback models used in the simulation for a given UVB.

\section*{Acknowledgments}
{
We acknowledge the use of High-performance computing facilities PERSEUS and PEGASUS at IUCAA. We are grateful to
K. Subramanian, S. Muzahid,  Vikram Khaire, and A. Mohapatra for insightful discussions regarding this work.
The Sherwood simulations were performed using the Curie supercomputer at the Tre Grand Centre de Calcul (TGCC), and the DiRAC Data Analytic system at the University of Cambridge, operated by the University of Cambridge High-Performance Computing Service on behalf of the STFC DiRAC HPC Facility
(www.dirac.ac.uk) . This was funded by BIS National E-infrastructure capital grant (ST/K001590/1), STFC capital grants ST/H008861/1 and ST/H00887X/1, and STFC DiRAC Operations grant ST/K00333X/1. DiRAC is part of the National E-Infrastructure. 
}

\section*{Data Availability}
The data underlying this article are available in the article and in its online supplementary material.




\bibliographystyle{mnras}
\bibliography{main} 

\bsp	
\label{lastpage}

\end{document}